\documentclass[12pt, preprint]{aastex}
\usepackage{lineno}
\usepackage{rotating}
\linenumbers

\usepackage{natbib}
\newcommand{\rxj}{RX~J1713.7$-$3946}

\shorttitle{Observations of \rxj\ with the Fermi LAT}

\begin{document}
\title{Observations of the young Supernova remnant \rxj\ with the
  {\emph{Fermi}} Large Area Telescope}

\author{
A.~A.~Abdo\altaffilmark{2}, 
M.~Ackermann\altaffilmark{3,1}, 
M.~Ajello\altaffilmark{3}, 
A.~Allafort\altaffilmark{3}, 
L.~Baldini\altaffilmark{4}, 
J.~Ballet\altaffilmark{5}, 
G.~Barbiellini\altaffilmark{6,7}, 
M.~G.~Baring\altaffilmark{8}, 
D.~Bastieri\altaffilmark{9,10}, 
R.~Bellazzini\altaffilmark{4}
% ...
}
\altaffiltext{1}{Corresponding authors: M.~Ackermann, markusa@slac.stanford.edu; S.~Funk, funk@slac.stanford.edu; Y.~Uchiyama, uchiyama@slac.stanford.edu.}
\altaffiltext{2}{Center for Earth Observing and Space Research, College of Science, George Mason University, Fairfax, VA 22030, resident at Naval Research Laboratory, Washington, DC 20375}
\altaffiltext{3}{W. W. Hansen Experimental Physics Laboratory, Kavli Institute for Particle Astrophysics and Cosmology, Department of Physics and SLAC National Accelerator Laboratory, Stanford University, Stanford, CA 94305, USA}
\altaffiltext{4}{Istituto Nazionale di Fisica Nucleare, Sezione di Pisa, I-56127 Pisa, Italy}
\altaffiltext{5}{Laboratoire AIM, CEA-IRFU/CNRS/Universit\'e Paris Diderot, Service d'Astrophysique, CEA Saclay, 91191 Gif sur Yvette, France}
\altaffiltext{6}{Istituto Nazionale di Fisica Nucleare, Sezione di Trieste, I-34127 Trieste, Italy}
\altaffiltext{7}{Dipartimento di Fisica, Universit\`a di Trieste, I-34127 Trieste, Italy}
\altaffiltext{8}{Rice University, Department of Physics and Astronomy, MS-108, P. O. Box 1892, Houston, TX 77251}
\altaffiltext{9}{Istituto Nazionale di Fisica Nucleare, Sezione di Padova, I-35131 Padova, Italy}
\altaffiltext{10}{Dipartimento di Fisica ``G. Galilei", Universit\`a di Padova, I-35131 Padova, Italy}
% ...

\begin{abstract}
  We present observations of the young Supernova remnant (SNR)
  RX\,J1713.7$-$3946 with the {\emph{Fermi}} Large Area Telescope
  (LAT). We clearly detect a source positionally coincident with the
  SNR. The source is extended with a best-fit extension of
  $0.55^{\circ} \pm 0.04^{\circ}$ matching the size of the
  non-thermal X-ray and TeV gamma-ray emission from the remnant. The
  positional coincidence and the matching extended emission allows us
  to identify the LAT source with the supernova remnant
  RX\,J1713.7$-$3946. The spectrum of the source can be described by a
  very hard power-law with a photon index of $\Gamma = 1.5 \pm 0.1$
  that coincides in normalization with the steeper H.E.S.S.-detected
  gamma-ray spectrum at higher energies. The broadband gamma-ray
  emission is consistent with a leptonic origin as the dominant
  mechanism for the gamma-ray emission.
\end{abstract}


\section{Introduction}

Gamma-ray observations of shell-type supernova remnants (SNRs) hold
great promise to help understanding the acceleration of cosmic rays
(CRs). These particles -- arriving at Earth mostly in the form of
protons -- are thought to be accelerated by a mechanism called
{\emph{diffusive shock}} acceleration~\citep{Bell-1,
  BlandfordOstriker, Jones, MalkovDrury} in the shocks of supernova
explosions up to energies around the ``knee'' in the spectrum of
cosmic rays ($\sim 10^{15}$~eV). In particular, X-ray and TeV
gamma-ray observations of young SNRs such as Cas~A
~\citep{HwangLaming, CasAChandra, MAGIC:casA, Fermi:casA}, or \rxj\
\citep{Koyama1713, UchiyamaASCA, HESS:rxj1713p2,HESS:rxj1713p3} have
confirmed the existence of relativistic particles in the shock
waves. Young SNRs are preferred targets for seeing particle
acceleration at work since in these objects the shocks are still
strong and actively accelerating particles to the highest
energies. Gamma-ray instruments have the angular resolution to
spatially resolve some of the closer SNRs.

\rxj\ (also known as G347.3$-$0.5) is a young ``historical'' remnant
suggested to be associated with the appearance of a guest star in the
constellation of Scorpius in AD393 by~\citet{Wang}. \rxj\ is located
in the Galactic plane (at $l=347.3^{\circ}$, $b=-0.5^{\circ}$) and was
discovered in soft X-rays in 1996 in the ROSAT all-sky
survey~\citep{Pfeffermann}. At a suggested distance of
1~kpc~\citep{Koyama1713, FukuiUchiyama, CassamXMM} with angular
diameter $\sim 65^{\prime} \times 55^{\prime}$, the size of the shell
is $\sim 20$~pc. Its properties are strikingly dominated by
non-thermal activity. Its X-ray emission shows a featureless spectrum
interpreted to be completely dominated by X-ray synchrotron emission
from ultra-relativistic electrons~\citep{Koyama1713, Slane1713,
  Taka2008}. The thermal X-ray emission as well as the radio emission
are extremely faint~\citep{Lazendic}. Detailed X-ray observations with
{\emph{Chandra}} and {\emph{XMM-Newton}} unveiled a complex structure
of filaments and knots in the shell of the SNR -- in particular in
the western part~\citep{Uchiyama1713, Lazendic, CassamXMM,
  AceroXMM}. A recent study with the {\emph{Suzaku}} satellite
extended the X-ray spectrum to $\sim 40$~keV, a measurement that
enabled the determination of the parent electron spectrum in the energy
range where the spectrum cuts off~\citep{Taka2008}.

\rxj\ is the first SNR for which TeV gamma-ray emission was
clearly detected emerging from the shell. H.E.S.S.\ measurements
provided the first-ever resolved gamma-ray emission at TeV
energies. The TeV emission closely matches the non-thermal X-ray
emission as demonstrated by~\citet{HESS:rxj1713p2}. The energy
spectrum of \rxj\ has been measured up to $\sim 100$~TeV, clearly
demonstrating particle acceleration to beyond these energies in the
shell of the SNR.

While the non-thermal X-rays detected in the shells of young SNRs are
clearly generated through synchrotron emission by ultra-relativistic
electrons~\citep{Koyama1713}, the picture of the particle population
radiating the gamma rays is not so clear.  The main argument revolves
around two main emission mechanisms~\citep{HESS:rxj1713p2, Katz,
  BerezhkoVoelk, Porter, Ellison, Morlino}, but so far, conclusive
evidence for either possibility is still missing. One scenario
suggests a connection of the TeV gamma-ray emission with accelerated
protons (CRs) by invoking the interaction of these protons with
interstellar material generating neutral pions ($\pi^0$s) which in
turn decay into gamma rays.  A second competing channel exists in the
inverse Compton scattering of the photon fields in the surroundings of
the SNR by the same relativistic electrons that generate the
synchrotron X-ray emission. This channel naturally accounts for
    the close resemblance between the X-ray and the TeV gamma-ray
    images. Several ways have been suggested to distinguish between
these two scenarios \citep[see e.g.][]{Morlino} but one of the most
promising seems to be the broadband modeling of the spectral energy
distribution (SED). In this regard, observations of young SNRs with
the LAT on board the {\emph{Fermi}} Gamma-Ray Space Telescope
({\emph{Fermi}}) are of particular importance since the LAT detects
gamma rays in the energy range that bridges sensitive measurements
with X-ray satellites such as {\emph{Chandra}} and {\emph{XMM-Newton}}
and TeV gamma-ray telescopes such as H.E.S.S., VERITAS or MAGIC.

\section{Observation and Analysis}

The \emph{Fermi}-LAT is a pair-conversion gamma-ray telescope with a
precision tracker and calorimeter, each consisting of a $4\times 4$
array of 16 modules, a segmented anti-coincidence detector (ACD) that
covers the tracker array, and a programmable trigger and data
acquisition system. The incoming gamma rays produce electron-positron
pairs in the tracker subsystem, which allow a reconstruction of the
directions of the primary gamma rays using the information provided by
the 36 layers of silicon strip detectors in the tracker. The energy of
the incoming gamma ray is determined from the energy deposited by the
electromagnetic showers in the segmented CsI calorimeter.  The ACD
subsystem is used as a veto against the great majority of cosmic rays
that trigger the LAT. The energy range of the LAT is 20 MeV to $> 300$
GeV with an angular resolution for events converting in the front part
of the detector of approximately
3.5$^\circ$ at 100 MeV, improving to about 0.1$^\circ$ at 10 GeV
(defined as the 68\% containment radius of the LAT point-spread
function or PSF).  Full details on the instrument and the on-board and
ground data processing are given in \citep{LATPaper}.

The LAT normally operates in a scanning mode (the ``sky survey'' mode)
that covers the whole sky every two orbits ($\sim$3 h). We use data
taken in this mode from the commencement of scientific operations on
2008 August 4 to 2010 August 4. The data were prepared and analyzed
using the LAT Science Tools package (v9r16p1), which is available from
the {\emph{Fermi}} Science Support Center
\footnote{http://fermi.gsfc.nasa.gov/ssc/}. Only events satisfying the
standard low-background event selection (the so-called ``Diffuse''
class events) and coming from zenith angles $< 105^{\circ}$ \citep[to
greatly reduce the contribution by Earth albedo gamma rays,
see][]{FermiAlbedo} were used in the present analysis.  We use all
gamma rays with energy $> 500$~MeV within a $20^{\circ} \times
20^{\circ}$ region of interest (ROI) centered at the nominal position
of \rxj\ ($\alpha=258.39^{\circ}$, $\delta=39.76^{\circ}$). We
    chose a lower bound of 500 MeV for this analysis for two reasons:
    Due to the relative hardness of the spectrum of \rxj\ compared to
    the Galactic diffuse background, photons with energies below 500
    MeV are not effective in constraining morphology or spectral shape
    of the source. Additionally, the broadening of the PSF at
    low energies might lead to systematic problems of confusion with
    neighboring sources in this densely populated region of the sky.
To further reduce the effect of Earth albedo backgrounds, any time
intervals when the Earth was appreciably in the field of view
(specifically, when the center of the field of view was more than
52$^{\circ}$ from the zenith) as well as any time intervals when parts
of the ROI were observed at zenith angles $> 105^{\circ}$ were also
excluded from the analysis.  The spectral analysis was performed based
on the P6v3 version of post-launch instrument response functions
(IRFs) which take into account pile-up and accidental coincidence
effects in the detector subsystems \citep{Rando2009}. The binned
maximum-likelihood mode of \emph{gtlike}, which is part of the
ScienceTools, was used to determine the intensities and spectral
parameters presented in this paper.

\subsection{Background sources}

\begin{figure}[htp!!]
  \centering
 \includegraphics[width=1.1\textwidth]{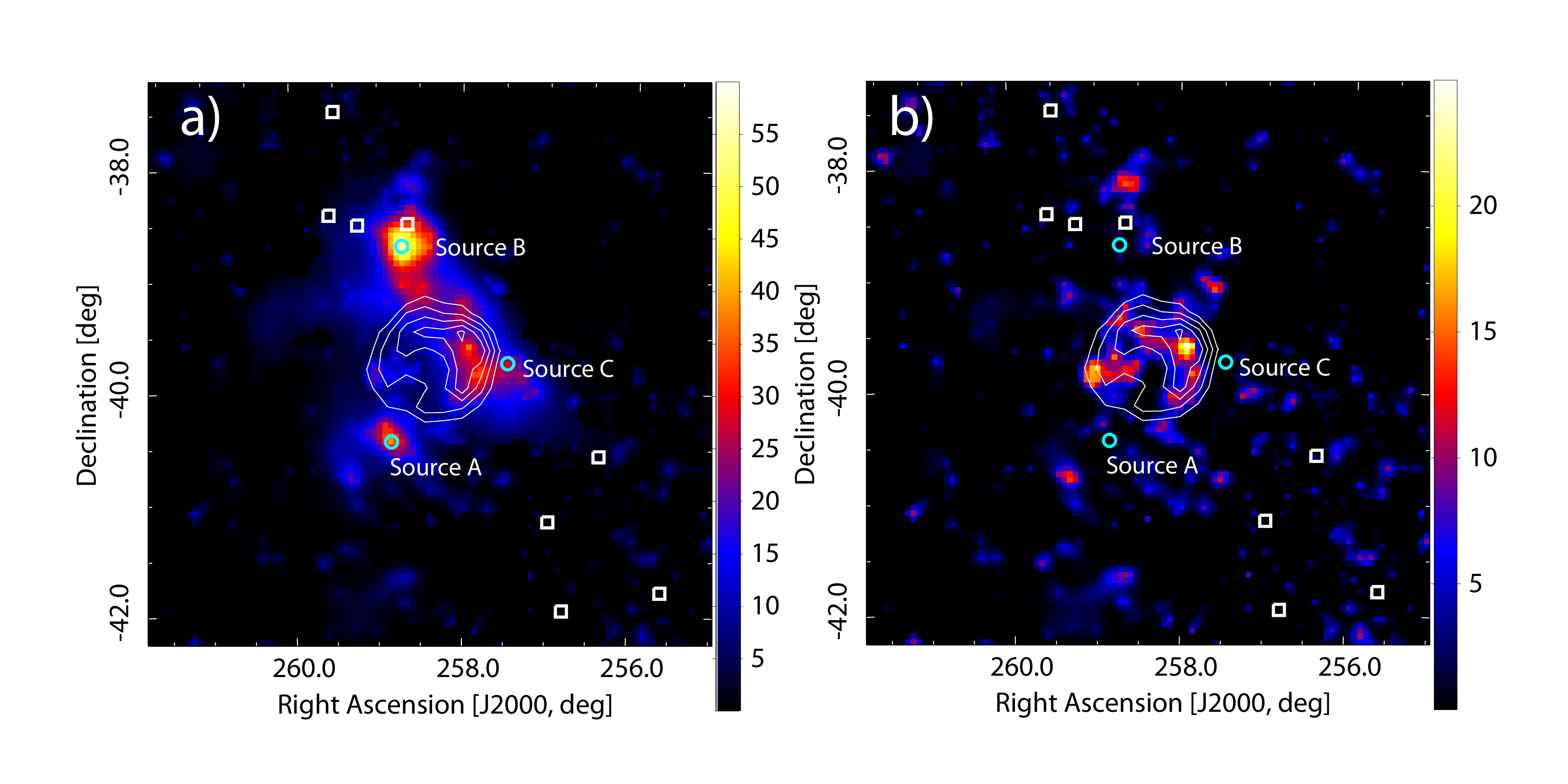} 
 \caption{
   {\bf Panel (a):} Map of the test statistic (TS) for a point source
   in the region around \rxj\ obtained in a maximum likelihood fit accounting for
   the background diffuse emission and 1FGL catalog sources.
   Only events above 500~MeV have been used in this
   analysis. H.E.S.S.\ TeV emission contours are shown in white~\citep{HESS:rxj1713p3}. 
   Rectangles indicate the positions of 1FGL sources in our background model,
   Several TS peaks outside the SNR shell are visible. The 3 peaks marked 
   by circles are added as additional sources to our background model (see text).
   {\bf Panel (b):} Same map as panel (a), but with the 3 additional sources now considered
   in the background model.}
  \label{fig:tsmap}
\end{figure}

We adopt a background model for the region which includes components
describing the diffuse Galactic and isotropic gamma-ray emission
\footnote{The LAT standard diffuse emission models
  ({\emph{gll\_iem\_v02.fits}} and {\emph{isotropic\_iem\_v02.fits}}), available at\\
  http://fermi.gsfc.nasa.gov/ssc/data/access/lat/BackgroundModels.html}. It
also includes all point sources within our ROI which are identified in
the 1FGL catalog \citep{LATcat:2010} except 1FGL 1711.7$-$3944c which
is spatially coincident with \rxj. All 1FGL sources are modeled with a
power-law spectrum using the flux and spectral index values obtained
from the catalog. Exceptions are the known pulsars in the ROI which we
model with a power-law with exponential cutoff spectral model. As the
parameters for this spectral model cannot be obtained from the 1FGL
catalog, we keep the flux, spectral index and cutoff energy of the
known pulsars as free parameters in the maximum likelihood fits of the
ROI.  Figure \ref{fig:tsmap} shows two maps of the point-source
detection significance, evaluated at each point in the map (TS map)
for the region around \rxj\ using photons with energies
$>$~500~MeV. The flux of the source is not permitted to be negative,
this is why negative fluctuations are not visible. The detection
significance is shown in terms of the test statistic (TS) of the
likelihood fit. The TS value is defined as TS=$ 2 ( \ln L_1/L_0 )$,
proportional to the logarithm of the likelihood ratio between a
point-source hypothesis ($L_1$) and the null hypothesis of pure
background ($L_0$) \citep{Mattox}. The significance contours of the
TeV emission observed from the SNR by the H.E.S.S.\ telescope array
\citep{HESS:rxj1713p2} are overlaid on the maps. Panel (a) shows the
TS map characterizing the excess emission found in the region around
\rxj\ over our background model described above. A significant TS
value is found within the spatial extent of the SNR but also in
several regions outside of its shell.

We identify three regions of excess gamma-ray emission which are
likely not associated with the SNR but belong to background sources
not recognized in the first {\emph{Fermi}} catalog (1FGL). Due to the
longer integration time of our analysis (24 months vs. 11 months in
the catalog) and the corresponding improved sensitivity, the
appearance of additional sources in our region of interest is
expected.  We simply denote these sources with the identifier
{\emph{A, B, C}}. The source positions are shown in Figure
\ref{fig:tsmap} and given in Table \ref{table:plawfit}.  The location
of source {\emph{A}} is consistent with a weak radio source
\citep{Lazendic}.  It is further identified in an internal update of
the {\emph{Fermi}} LAT catalog using 24 months of data.  Source
{\emph{B}} is only 11$^{\prime}$ from the catalog source
1FGL~J1714.5$-$3830c and could be an artifact caused by unmodeled
emission from 1FGL~J1714.5$-$3830c if this source were spatially
extended as has been tentatively suggested
by~\citet{CastroSlane}. 1FGL~J1714.5$-$3830c is modeled as a point
source in the 1FGL catalog.  However, the catalog source is spatially
coincident with the SNR CTB~37A which has an extent in radio of $\sim
15^{\prime}$ \citep{green04:SNRs}.  A detailed study of the morphology
of this source is in progress but beyond the scope of this publication
as the exact morphology of the CTB~37A source does not significantly
affect the spectral analysis of \rxj. For simplicity we just assume
the emission from this region to be described by two independent point
sources, 1FGL~J1714.5$-$3830c and source {\emph{B}}. The third
additional background source {\emph{C}} shown in Figure
\ref{fig:tsmap} may be associated with RX\,J1713.7$-$3946. It is very
close to \rxj, located about 35$^{\prime}$ from the center of the SNR.
However, it is spatially consistent with a local enhancement of
molecular gas, observed via the radio emission from the CO
($J$=1$\rightarrow$0) transition \citep{Dame:12CO}. Furthermore, we
will show below (see Table \ref{table:plawfit}) that in a combined
likelihood analysis of the spectra of \rxj\ and the surrounding
background sources the emission from source {\emph{C}} is considerably
softer than the gamma-ray emission from the SNR. In fact, both the
spectral index and the intensity of the source are consistent with
expectations of gamma-ray emission from a small cloud of molecular
gas. Nevertheless, we cannot reject the possibility that at least part
of the emission attributed to the additional background source
{\emph{C}} is originating from the SNR shell.  While we consider
source {\emph{C}} an independent point source in our standard
background model of the ROI, we repeat the spectral analysis with a
model without this source and account for the difference in our
estimation of systematic uncertainties. Panel (b) in Figure
\ref{fig:tsmap} shows the detection significance map for the region
around \rxj\ (E~$> 500$~MeV) with our standard background model
accounted for.  A comparison with the significance contours from
H.E.S.S.\ suggests a spatially extended emission from the shell of the
SNR rather than a single point source.

\subsection{Centroid and Angular Extent}

We study the morphology of the emission associated with
RX\,J1713.7$-$3946 with a series of maximum likelihood fits, comparing
the TS value for different hypotheses about the shape and extent of
the source. We fitted the extension (and position) of the gamma-ray
emission with a disk of varying radius. The emission is found to be
significantly extended; the best-fit position (RA, Dec =
$258.50^{\circ} \pm 0.04^{\circ}_{\mathrm{stat}}, -39.91^{\circ} \pm
0.05^{\circ}_{\mathrm{stat}}$) is consistent with the center of the SNR
  within $0.2^{\circ}$ and the best-fit radius is $0.55^{\circ} \pm
  0.04^{\circ}$. This size is consistent with that of the X-ray SNR
  given in~\citet{green04:SNRs} as $1.1^{\circ} \times 0.9^{\circ}$ in
  diameter. To confirm these fits, we test a single point source at
  the location of the highest excess in the TS map within the shell of
  the SNR. We further test a spatially extended source defined by the
  shape of the H.E.S.S.\ significance contours of \rxj\ and an
  extended source as a uniform disk of 0.55$^{\circ}$ radius. Finally,
  we consider two and three independent point sources within the shell
  of the SNR located at the most prominent peaks in the TS map. A
  power-law spectrum with integrated flux (between 1 and 300~GeV) and
  spectral index as free parameters is assumed for each of the
  hypotheses. The detailed setup of the likelihood fit is identical to
  the one used for the spectral analysis and described with that
  analysis (Section \ref{sec::spectrum}). Table \ref{table:morpho}
  shows the flux, and spectral index of the tested shape and its TS
  value in comparison to the background model. The TS values are
  suggestive of extended gamma-ray emission from RX\,J1713.7$-$3946.
  The H.E.S.S.\ significance map as well as the uniform disk have a
  difference in TS of $\Delta {\mathrm{TS}} = 61$ or
  $58$~(H.E.S.S./Disk) relative to a single point source and a $\Delta
  {\mathrm{TS}} = 43$ or $40$~(H.E.S.S./Disk) relative to a set of 3
  point sources within the shell of \rxj. However, the TS value in a
  comparison to the background model for both the H.E.S.S.\
  significance map ($TS=77$) and the uniform disk ($TS=79$) are almost
  identical, demonstrating that we are not sensitive to the detailed
  shape of the emission region.  For the models of RX\,J1713.7$-$3946
  considered, the TS value is expected to follow a
  $\chi^{2}$-distribution with two degrees of freedom in the case that
  no source is present \citep{Mattox} and therefore can be converted
  to a detection significance of $\sim 8.5\sigma$ for both the
  H.E.S.S.\ template and the uniform disk model. The positional and
  the angular-size coincidence with the X-ray and TeV gamma-ray
  emission strongly favors an identification of the LAT source with
  the SNR RX\,J1713.7$-$3946.

\begin{figure}[htp!!]
  \centering
 \includegraphics[width=1.1\textwidth]{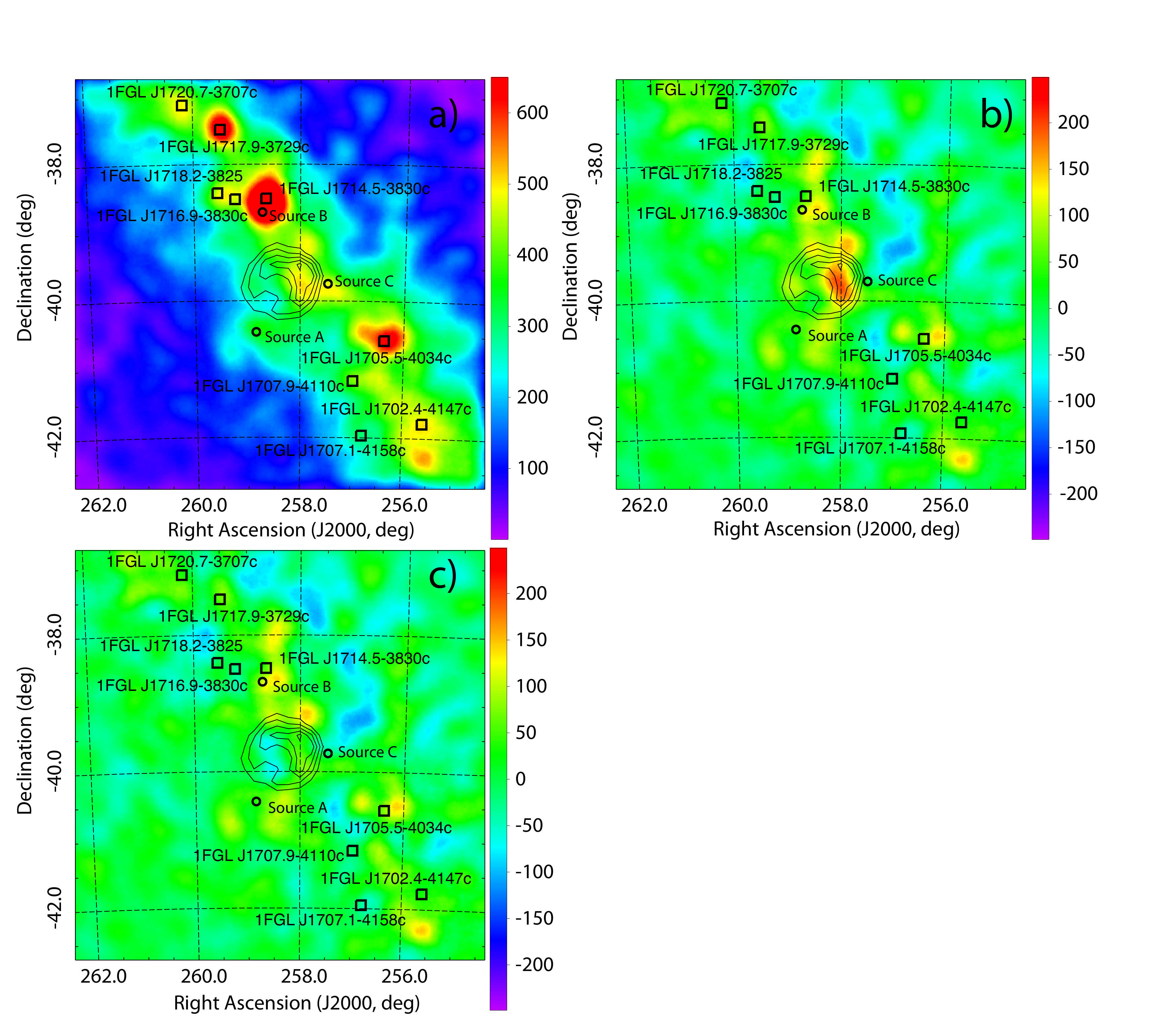} 
 \caption{ {\bf Panel (a):} Counts/sq.~deg. observed by the {\emph{Fermi}} LAT
   above 3 GeV in the region around \rxj.  The map is smoothed with a
   0.3$^\circ$-wide Gaussian kernel corresponding to the width of the
   LAT PSF at 3 GeV. H.E.S.S.\ TeV emission contours are shown in
   black~\citep{HESS:rxj1713p3}.  Rectangles indicate the positions of
   1FGL sources. Circles indicate the additional sources considered in
   our background model.  {\bf Panel (b):} Residual counts after the
   subtraction of the counts attributed to the background model.  {\bf
     Panel (c):} Residual counts after the subtraction of the counts
   attributed to the background model and to \rxj.}
  \label{fig:countmap}
\end{figure}

Fig. \ref{fig:countmap} shows a series of LAT gamma-ray counts maps of
the sky surrounding \rxj.  We choose an energy threshold of 3 GeV for
these maps, higher than the analysis threshold of 500 MeV, to enhance
their resolution. The counts maps are smoothed with an 0.3$^{\circ}$
wide Gaussian kernel. This width corresponds to the size of the LAT
PSF at 3 GeV (the 39\% containment radius of a 2-D gaussian), averaged
over front and back conversions and over all incident
angles. Locations of 1FGL catalog sources in the region are marked by
squares. Our additional background sources are denoted by circles and
labeled.  The black lines again display the contours of the
H.E.S.S. significance map of \rxj.  Panel~(a) shows all counts in the
region. The emission coinciding with \rxj\ is faint; the counts map is
dominated by the Galactic diffuse emission as well as emission from
1FGL~J1714.5$-$3830c and 1FGL~J1705.5$-$4034c. Panel~(b) shows a
residual counts map after subtraction of our background model. On this
panel a clear excess within the shell of \rxj\ is visible. Panel~(c)
finally shows the residual counts after subtraction of our background
model as well as the emission from \rxj\ (using the
H.E.S.S. significance map as the template for the spatial
extension). The residual counts are consistent with the expected
statistical fluctuations, i.e the region around the SNR is well
described by our model.

\subsection{Spectral Analysis}
\label{sec::spectrum}

We adopt the spatial extension model based on the
H.E.S.S. significance map as the default model for the analysis of the
spectrum of \rxj. As discussed in the previous section, the LAT is not
able to distinguish between the two extended source models that we
tested. Therefore, we compare the obtained spectrum from the default
model to the results derived from a uniform disk source model and
include the difference in the systematic uncertainty of the spectrum.
In the first step of the spectral analysis we perform a maximum
likelihood fit of the spectrum of \rxj\ in the energy range between
500~MeV and 400~GeV using a power-law spectral model with integral
flux and spectral index as free parameters.  To accurately account for
correlations between close-by sources we also allow the integral
fluxes and spectral indices of the nearby 1FGL and sources {\emph{A,
    B, C}} ($<$~3$^{\circ}$ from the center of the ROI) to be free for
the likelihood maximization, as well as the spectral parameters of
identified LAT pulsars, instead of fixing them to the 1FGL catalog
values.  We redetermine in our fit the normalization of the Galactic
diffuse emission model, the index of an energy dependent (power-law)
multiplicative correction factor to it, and the normalization of the
isotropic component. This accounts for localized variations
in the spectrum of the diffuse emission in the fit which are not
considered in the global model.

\begin{figure}[htp!!]
  \centering
 \includegraphics[width=0.6\textwidth]{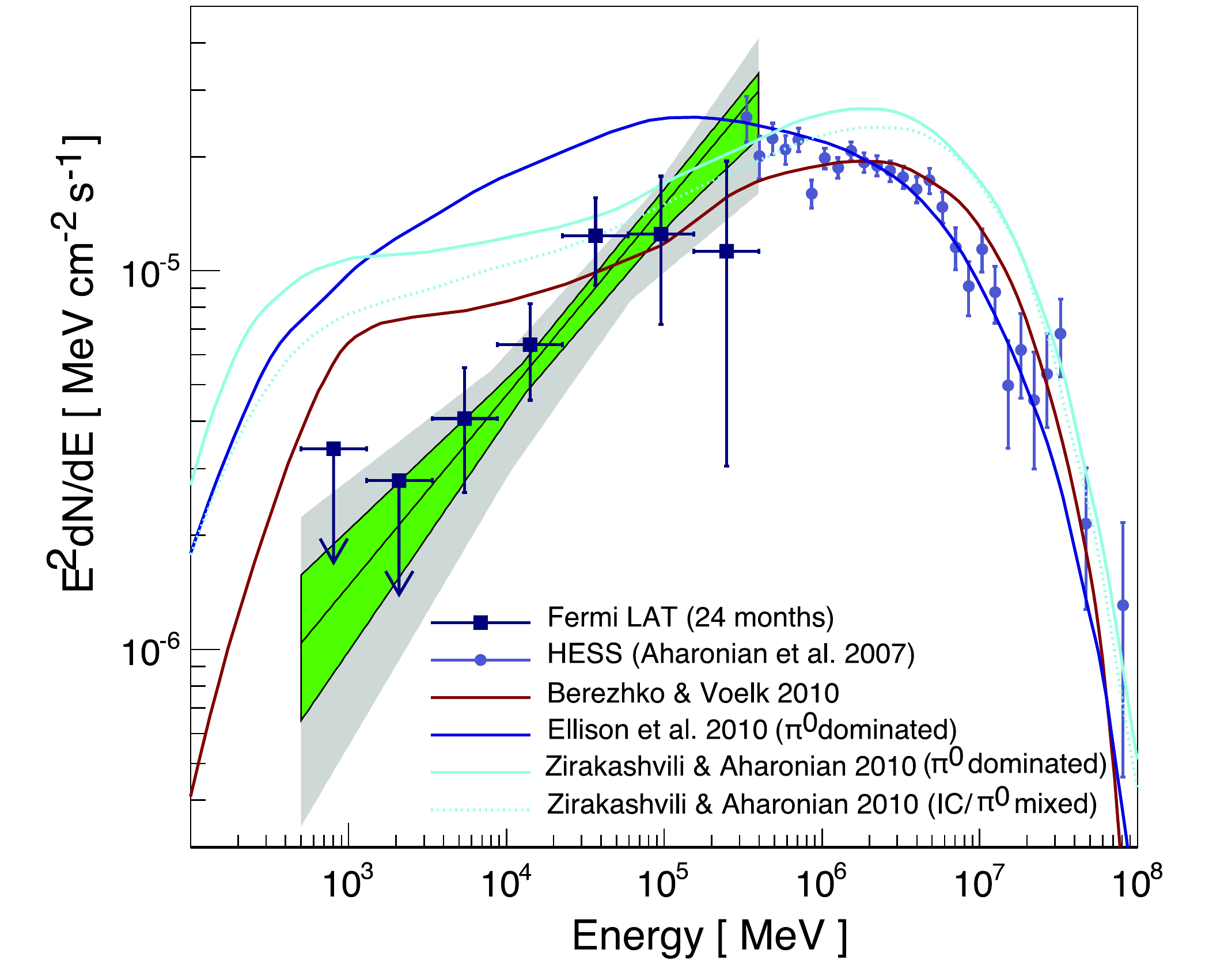}
 \includegraphics[width=0.6\textwidth]{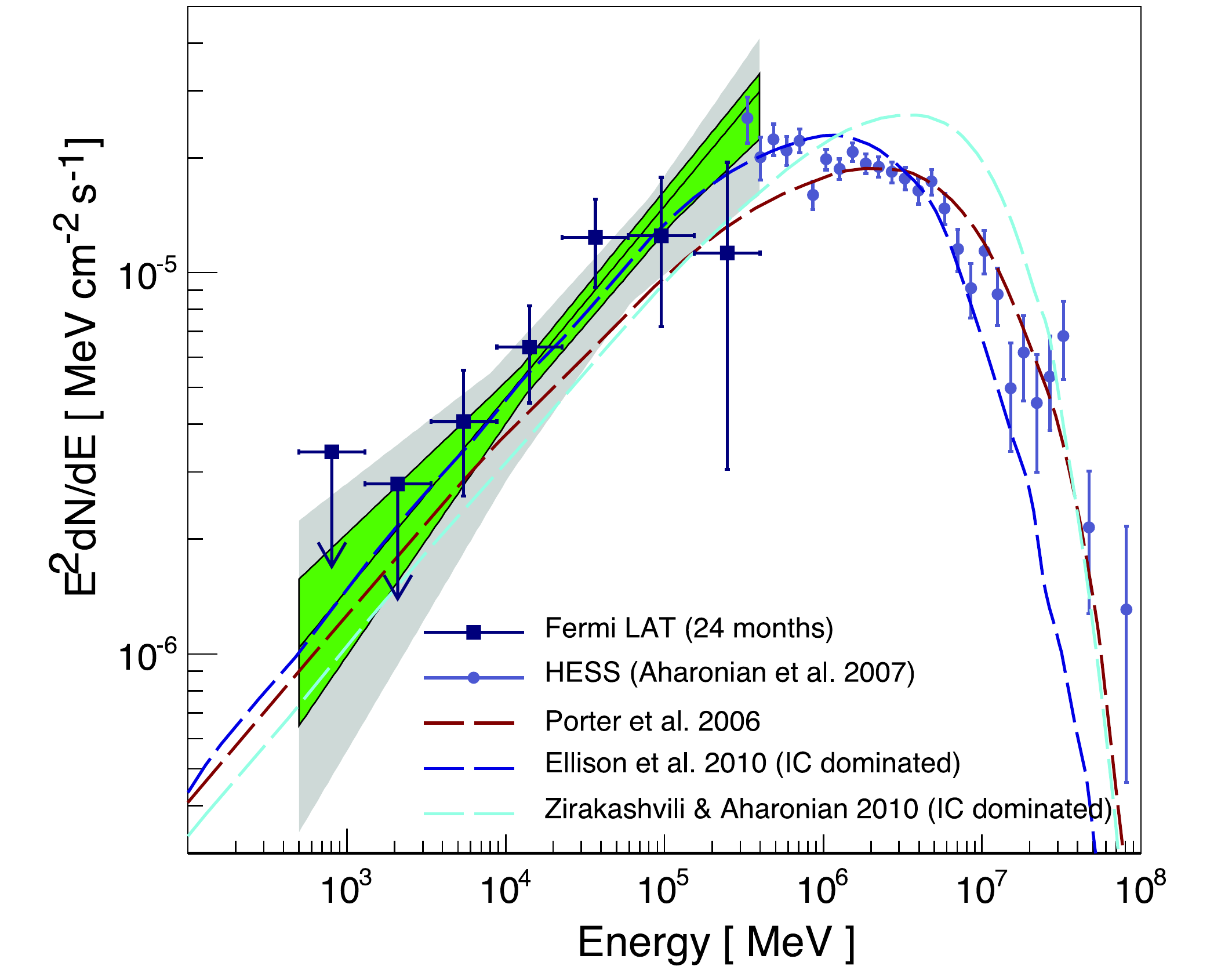}
 \caption{{\footnotesize{Energy spectrum of \rxj\ in gamma rays. Shown
       is the {\emph{Fermi}}-LAT detected emission in combination with
       the energy spectrum detected by
       H.E.S.S.~\citep{HESS:rxj1713p3}.  The green region shows the
       uncertainty band obtained from our maximum likelihood fit of
       the spectrum of \rxj\ assuming a power-law between 500~MeV and
       400~GeV for the default model of the region. The gray region
       depicts the systematic uncertainty of this fit obtained by
       variation of the background and source models. The black error
       bars correspond to independent fits of the flux of \rxj\ in the
       respective energy bands. Upper limits are set at 95\%
       confidence level.  Also shown are curves that cover the range
       of models proposed for this object. These models have been
       generated to match the TeV emission and pre-date the LAT
       detection. The top panel features predictions assuming that the
       gamma-ray emission predominately originates
       from 
       the interaction of protons with interstellar gas (brown:
       \citet{BerezhkoVoelk}, blue: \citet{Ellison}, cyan
       (solid/dashed): \citet{Zirakashvili}).  The bottom panel
       features models where the bulk of the gamma-ray emission arises
       from interactions of electrons with the interstellar radiation
       field (leptonic models).  (brown: \citet{Porter}, blue:
       \citet{Ellison}, cyan: \citet{Zirakashvili}).  See text for a
       qualitative discussion of these models.}}  }
  \label{fig:sed}
\end{figure}

For the Galactic diffuse emission, we find a normalization factor of
0.93~$\pm$~0.01 in our region of interest and a spectral correction
factor index of 0.019~$\pm$~0.002 (the positive sign corresponds to a
spectrum that is harder than in the model). The normalization factor
for the isotropic component is 1.17~$\pm$~0.05. These factors
    demonstrate the good agreement of the local brightness and
    spectrum of the diffuse gamma-ray emission with the global diffuse
    emission model. Table \ref{table:plawfit} summarizes the source
parameters obtained as results from this fit.  The table includes the
spectral parameters and the TS values of all fitted sources.  The flux
above 1~GeV obtained for \rxj\ with our default background model is
$F_{1000}$~$=($2.8~$\pm$~0.7)$\times 10^{-9}$~cm$^{-2}$~s$^{-1}$ and
the spectral index is $\Gamma$~$=$~1.50~$\pm$~0.11.  Figure
\ref{fig:sed} shows the uncertainty band obtained from this fit.

In a second step we perform a maximum likelihood fit of the flux of
\rxj\ in 7 independent logarithmically spaced energy bands from 500
MeV to 400 GeV (using the spectral model and parameters obtained in
the previous fit) to obtain a spectral energy distribution (SED) for
the SNR.  The resulting SED is displayed in Figure \ref{fig:sed} as
black error bars. We require a test statistic value of TS$\geq$4 in
each band to draw a data point corresponding to a 2 $\sigma$
    detection significance. This criterion is not fulfilled for the
lowest two energy bands 500~MeV--1.3~GeV and 1.3~GeV--3.4~GeV and
accordingly we show 95\% flux upper limits for these bands.
   
In a final step we estimate the systematic uncertainty on the obtained
spectral parameters by repeating the maximum likelihood analysis for
several variations of our default model. Specifically, we varied the
source shape template, the background sources, and the model of the
Galactic diffuse emission.  The spectral analysis was performed: a)
with the uniform disk shape replacing the H.E.S.S. significance map
template; b) with the closest background source {\emph{C}} removed
from the model (see also discussion above); c) using a preliminary
list of sources from the 2FGL catalog in development within the LAT
collaboration; d) replacing the standard diffuse emission model by a
refined model that is currently being evaluated in the collaboration
for source analysis for the 2FGL catalog (refined with 24 months of
data and with finer gas maps); e) replacing the standard diffuse model
by a model based on the GALPROP code \footnote{GALPROP is a software
  package for calculating the diffuse Galactic gamma-ray emission
  based on a model of cosmic-ray propagation in the Galaxy.  See
  http://galprop.stanford.edu/ for details and references} used in the
{\emph{Fermi}} LAT analysis of the isotropic diffuse emission.  The GALPROP
model is described in \cite{EGBpaper}. For e), i.e.\ the GALPROP-based
model, we considered the various components of the diffuse emission
model separately for which we then individually fit the normalizations
in our likelihood analysis. The components are gamma rays produced by
inverse Compton emission, gamma rays produced by interactions of CRs
with atomic and ionized interstellar gas and gamma rays produced in
the interactions of CRs with molecular gas. The model component
describing the gamma ray intensity from interactions with molecular
gas is further subdivided into seven ranges of Galactocentric distance
to accommodate localized variations of the CR and molecular gas
density along the line of sight which are not accounted for in the
model.

\begin{figure}[htp!!]
  \centering
 \includegraphics[width=0.6\textwidth]{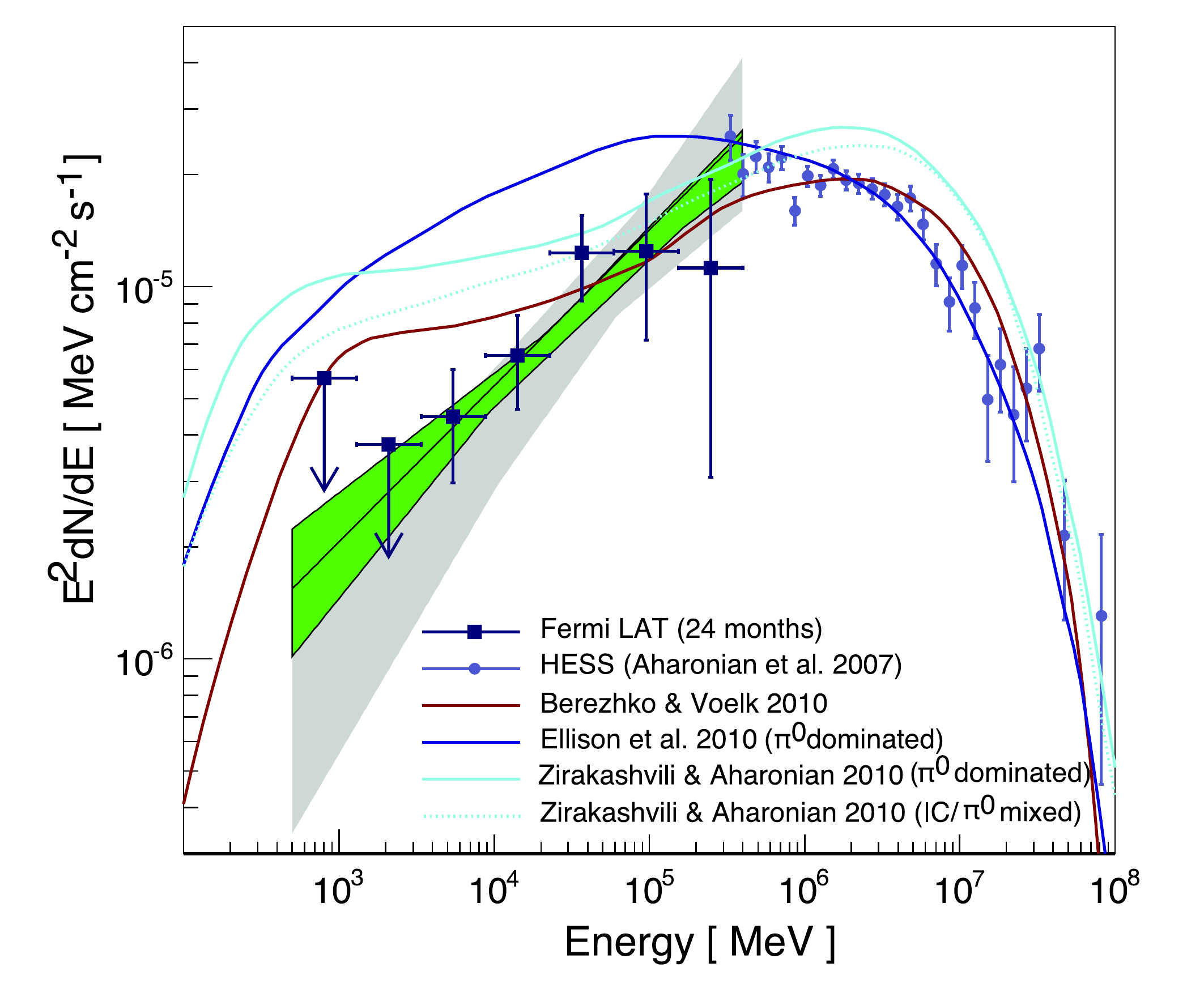}
 \includegraphics[width=0.6\textwidth]{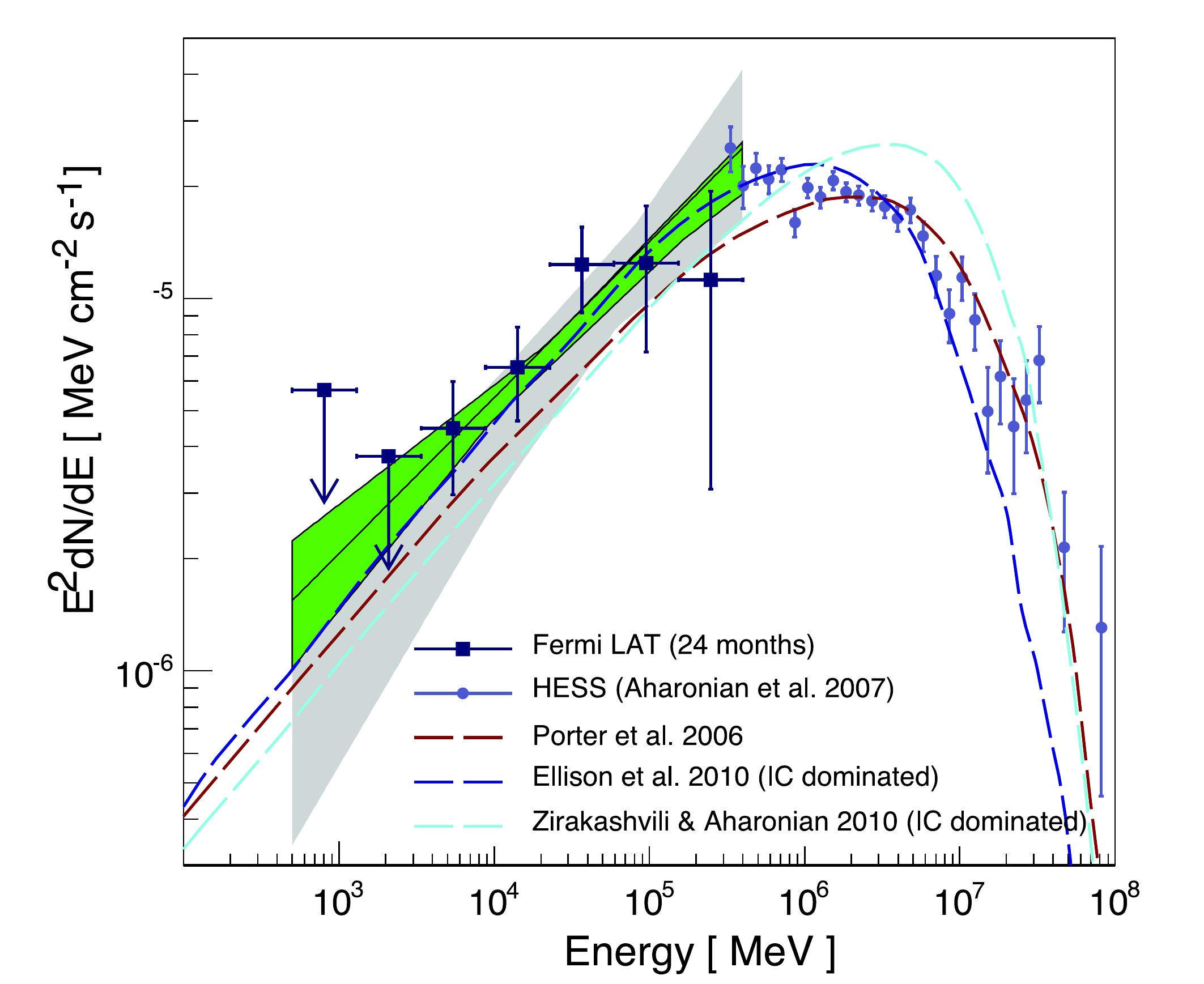}
 \caption{Same as Figure \ref{fig:sed} but featuring the 
source and background model which resulted in the softest spectrum for
\rxj\ instead of our default model.  }
  \label{fig:sedWC}
\end{figure}

The same model of the isotropic component was used for all model
variations a)--e). From the model variations a)~--~e) we obtain a
systematic uncertainty of +0.08/-0.10 for the spectral index of \rxj\
and a systematic uncertainty of (+0.6/-0.7)$\times
10^{-9}$~cm$^{-2}$~s$^{-1}$ for the flux above 1 GeV on top of the
statistical uncertainty. The systematic uncertainty of the derived
flux and spectral index related to the uncertainty in the LAT
effective area was evaluated separately. The uncertainty of the LAT
effective area -- estimated from observations of
Vela~\citep{FermiVelaPulsar} and the Earth Albedo~\citep{FermiAlbedo}
-- ranges from 10\% at 500 MeV to 20\% at $\geq$10 GeV. The impact on
the spectral parameters of \rxj\ is a systematic uncertainty of
$\pm$0.05 for the spectral index and a systematic uncertainty of
$\pm$0.4 for the flux above 1 GeV.  The gray band in Figure
\ref{fig:sed} displays the superposition of all uncertainty bands
obtained in our variations of the default model.  Figure
\ref{fig:sedWC} depicts the model variation (b) resulting in the
softest spectrum together with the fluxes in individual energy bands
(black error bars) derived for model (b) using the same procedure as
for the default model described above. The range of systematic
uncertainty is particularly important to consider for comparisons of
the spectrum to pion-decay dominated gamma-ray emission models which
are generally expected to be softer than inverse Compton dominated
gamma-ray emission models.

\section{Discussion}

The positional coincidence between the extended gamma-ray emission
detected by the {\emph{Fermi}}-LAT at the position of
RX~J1713.7$-$3946 strongly suggests a physical association between the
GeV gamma-ray emission and this young SNR. In addition, the region of
brightest LAT gamma-ray emission coincides with the northwestern part
of the SNR. From CO $(J = 1-0)$ observations~\citet{FukuiUchiyama} and
\citet{Moriguchi} suggested that this part of the SNR is undergoing
complex interactions between the supernova shock wave and a molecular
cloud. This part is also the brightest region in non-thermal X-rays
and in TeV gamma rays. The match between the locations of brightest
emission suggests that the GeV emission is also generated by the
population of relativistic particles responsible for the TeV gamma-ray
and non-thermal X-ray emission.

The origin of the TeV gamma-ray emission from RX~J1713.7$-$3946 has
been a matter of active debate \citep[see][and references
therein]{Zirakashvili}.  There are two competing processes potentially
responsible for the shell-like TeV gamma-ray emission from
RX~J1713.7$-$3946: Inverse Compton (IC) scattering on the cosmic
microwave background by relativistic electrons (leptonic model) and
$\pi^0$-decay gamma rays resulting mainly from inelastic collisions
between relativistic protons and ambient gas nuclei (hadronic model).
It is generally accepted that diffusive shock acceleration (DSA)
operates at supernova shocks producing high-energy protons and
electrons.  However, injection mechanisms of supra-thermal particles
are poorly known so that the current theory cannot tell us about the
number of relativistic protons and electrons produced at shocks. This
makes it difficult to reliably predict the levels of leptonic and
hadronic gamma-rays.

The lack of thermal X-ray lines provided a stringent constraint
on the gamma-ray production mechanisms.  The luminosity of hadronic
gamma-rays scales as $\bar{n}_{\rm H} W_p$, where $\bar{n}_{\rm H}$
denotes the gas density averaged over the emission volume (where
accelerated protons are assumed to be uniformly distributed), $W_p =
\xi E_{\rm SN}$ is a total energy content of accelerated protons, and
$E_{\rm SN} \sim 10^{51}\ \rm erg$ is the total kinetic energy
released by the SN explosion.  The lack of thermal X-ray emission in
SNR RX~J1713.7$-$3946 \citep{Slane1713,Taka2008} severely restricts
the gas density in the SNR to be small.
\citet{Ellison2010} have performed calculations of thermal X-ray
  emission from shocked plasma with non-equilibrium ionization in the
  case of uniform ambient density, following a hydrodynamic evolution
  with which non-linear DSA theory is coupled; they found that
the shocked gas densities required for consistency with the hadronic
model are $n_{\rm H} \la 0.2$ cm$^{-3}$. It should be noted
that, taking $E_{\rm SN} = 2\times 10^{51}\ \rm erg$, one needs $\xi
\sim 1$ (i.e., extremely efficient acceleration) for $\bar{n}_{\rm H}
= 0.1$ cm$^{-3}$ and $d=1\ \rm kpc$.  The extremely efficient
  (more efficient than usually assumed) transformation of the
  supernova kinetic energy into accelerated particles may lead to very
  low shocked gas temperature \citep{Drury09}, which in turn could
  change the density requirement.

The measurements of GeV gamma-ray emission obtained with the
{\emph{Fermi}}-LAT presented in this paper provide new, crucial
information about the particle population responsible for the
gamma-ray emission.  We have measured the gamma-ray spectrum of SNR
RX~J1713.7$-$3946 in the energy range from 500 MeV to 400 GeV and
found that the spectrum can be characterized by a hard power law with
photon index $\Gamma =1.5 \pm 0.1$(stat)$\pm 0.1$(sys), smoothly
connecting with the steeper TeV gamma-ray spectrum measured with
H.E.S.S.  Note that the measured gamma-ray spectrum of
RX~J1713.7$-$3946 now covers five orders of magnitude in energy,
unprecedented for SNRs.

The hard power-law shape in the \emph{Fermi}-LAT energy range with
photon index of $\Gamma = 1.5 \pm 0.1$ qualitatively agrees with the
expected IC spectrum of the leptonic model, as illustrated in both
Figures~\ref{fig:sed} and \ref{fig:sedWC}. If the leptonic model
  explains the gamma-ray spectrum, the \emph{Fermi}-LAT spectrum is
  emitted by a power-law part of the accelerated electrons, and
  therefore we can deduce the power-law index of electrons from the
  measured photon index. Using $\Gamma = 1.5 \pm 0.1$, we obtain $s_e
  = 2\Gamma -1 = 2.0 \pm 0.2$.  The energy flux ratio of the observed
synchrotron X-ray emission and the gamma-ray emission means that the
average magnetic field is weak, $B \simeq 10\, \mu$G
\citep{HESS:rxj1713p2,Porter, Ellison2010}. The maximum energy of
electrons is then $E_{e,\rm max} \sim 20\mbox{--}40$ TeV as determined
from the \emph{Suzaku} X-ray spectrum \citep{Taka2008}.  The presence
of synchrotron X-ray filaments varying on yearly timescales
\citep{Uchiyama2007}, if interpreted as being due to fast electron
acceleration and synchrotron cooling, requires $B \sim 0.1\mbox{--}1$
mG, which is difficult to reconcile with the weak average
field. Alternatively, the X-ray variability may be caused by
time-variable turbulent magnetic fields \citep{Bykov09} which require
a smaller magnetic field strength.  The filamentary structures and
variability in X-rays should be attributed to locally enhanced
magnetic fields in the case of the leptonic model~\citep{Pohl2005}.

As shown in Fig.~\ref{fig:sed}, several groups have previously
presented calculations of IC gamma-ray spectra. Detailed comparisons
between the observed total GeV--TeV spectrum and IC models show that
none of the previous IC models matches exactly with the data.  Some
additional complications would need to be introduced to realize a
better description of the gamma-ray data. For example, the shape of
the total IC spectrum could be modified if we add a second population
of electrons (or even multiple populations) which has a different
maximum energy \citep[see][]{Taka2008,Yamazaki2009}. Yet another way
of modifying the IC spectral shape is by invoking more-intense
interstellar radiation fields, though this would require substantial
increase in the photon density \citep[see][]{Taka2008}.

Even in the case of the leptonic model, it is important to constrain
the level of $\pi^0$-decay emission at GeV energies by allowing for a
hybrid (leptonic and hadronic) model of the GeV--TeV gamma-ray
spectrum.  For proton number index $s=2$ (assumed to be same as the
electron number index: see e.g. \citet{Baring1999} for a discussion of
why relativistic electron and ion indices should be very similar in
non-linear shocks), the GeV flux upper limit at 1 GeV corresponds to
$W_p < 0.3 \times 10^{51} (\bar{n}_{\rm H}/0.1\ \rm cm^{-3})^{-1}\ \rm
erg$ for $d = 1\ \rm kpc$, where $\bar{n}_{\rm H}$ denotes the
hydrogen number density of X-ray/gamma-ray emitting gas.  Therefore,
the leptonic model does not necessarily mean the proton content in
this SNR is unexpectedly small.

The GeV measurements with {\emph{Fermi}}-LAT do not agree with the
expected fluxes around 1 GeV in most hadronic models published so far
\citep[e.g.,][]{BV10}.  Given the current models of diffusive
  shock acceleration, we can discard the hadronic origin of the
  GeV--TeV gamma-ray emission.  The proton number index $s \sim 1.5$
inferred by the LAT spectrum is as small as the asymptotic index
of $s=1.5$ predicted by extremely efficient CR acceleration
\citep[][see also ~\citet{EllisonEichler} for early indications of
this limiting behavior]{Malkov99}.  Unless this asymptotic index
  is realized in the shock waves of RX~J1713.7$-$3946, the hard
  \emph{Fermi}-LAT spectrum cannot be ascribed to the $\pi^0$-decay
  emission.  However, such a proton energy distribution is not
  observed in the current models of efficient DSA \citep{Ellison2010}.

\section{Summary}

We have measured the GeV gamma-ray emission from RX~J1713.7$-$3946
with the {\emph{Fermi}}-LAT. The emission is extended and shows a size
that matches the TeV-detected gamma-ray emission from this SNR. The
gamma-ray spectrum for the SNR has been measured over more than 5
orders of magnitude combining {\emph{Fermi}}-LAT and H.E.S.S.\
observations. The spectral index in the {\emph{Fermi}}-LAT band is
very hard with a photon index of $1.5 \pm 0.1$ which is well in
agreement with emission scenarios in which the dominant source of
emission is Inverse Compton scattering of ambient photon fields of
relativistic electrons accelerated in the shock front. The dominance
of leptonic processes in explaining the gamma-ray emission does not
mean that no protons are accelerated in this SNR, but that the ambient
density is too low to produce a significant hadronic gamma-ray
signal. 
\rxj\ is the first remnant where the combination with H.E.S.S.\ data
yields spectroscopic measurements over more than 5 decade in energy
that, in contrast to many of the other LAT-detected remnants strongly
suggests a leptonic origin of the gamma-ray emission.
 
\acknowledgements{
The \textit{Fermi} LAT Collaboration acknowledges generous ongoing support
from a number of agencies and institutes that have supported both the
development and the operation of the LAT as well as scientific data analysis.
These include the National Aeronautics and Space Administration and the
Department of Energy in the United States, the Commissariat \`a l'Energie Atomique
and the Centre National de la Recherche Scientifique / Institut National de Physique
Nucl\'eaire et de Physique des Particules in France, the Agenzia Spaziale Italiana
and the Istituto Nazionale di Fisica Nucleare in Italy, the Ministry of Education,
Culture, Sports, Science and Technology (MEXT), High Energy Accelerator Research
Organization (KEK) and Japan Aerospace Exploration Agency (JAXA) in Japan, and
the K.~A.~Wallenberg Foundation, the Swedish Research Council and the
Swedish National Space Board in Sweden.

Additional support for science analysis during the operations phase is gratefully
acknowledged from the Istituto Nazionale di Astrofisica in Italy and
the Centre National d'\'Etudes Spatiales in France.}

 \begin{table}[htp!!]
\begin{minipage}{\linewidth}
  \renewcommand{\footnoterule}{}
  \begin{tabular}{l|ccccc}
    Source morphology & 
    Flux\footnote{{\footnotesize{E$>$1~GeV, in $10^{-9}$~cm$^{-2}$~s$^{-1}$}}} & 
    Photon index & 
    TS \footnote{{\footnotesize{TS value in
          comparison to a model with no source at the position of
          \rxj.}}} &
    R.~A. 2000&
    Dec.  \\
    \hline\hline
    Point source & 1.2~$\pm$~0.7 &  1.85~$\pm$~0.31 & 18 & 257.94$^{\circ}$ & $-$39.75$^{\circ}$  \\ 
    2 point sources & 0.5~$\pm$~0.5 & 1.68~$\pm$~0.41 & & 257.93$^{\circ}$ & $-$39.61$^{\circ}$ \\
    &  1.2~$\pm$~0.9 & 2.13~$\pm$~0.41 & 20 & 257.85$^{\circ}$ & $-$39.86$^{\circ}$ \\ 
    3 point sources & 0.5~$\pm$~0.5 & 1.69~$\pm$~0.41 & & 257.93$^{\circ}$  & $-$39.61$^{\circ}$ \\
    &  1.2~$\pm$~0.2 & 2.10~$\pm$~0.28 & & 257.85$^{\circ}$ & $-$39.86$^{\circ}$ \\ 
    & 0.4~$\pm$~0.3 &  1.61~$\pm$~0.31 & 32 & 259.00$^{\circ}$ & $-$39.81$^{\circ}$ \\ 
    Extended source (H.E.S.S.) 
    \footnote{{\footnotesize{H.E.S.S. significance map is used as a
          template for the intensity of the gamma-ray emission.}}} 
    & 2.8~$\pm$~0.7 & 1.50~$\pm$~0.11 & 77 & & \\
    Extended source (uniform disk)
    \footnote{{\footnotesize{A uniform disk with 0.55$^{\circ}$ radius
          is used as a template for the intensity of the gamma-ray
          emission. The specified coordinates correspond to the center
          of the disk. These parameters are the best-fit parameters
          when simultaneously fitting the position and the extension.}}} 
    & 3.2~$\pm$~0.7 & 1.49~$\pm$~0.10 & 79 & 258.50$^{\circ}$ & $-$39.91$^{\circ}$ \\
    \hline 
\end{tabular}
\end{minipage}
\caption{Results of the morphological analysis of the gamma-ray
  emission from \rxj. The integral flux between 1 and 300 GeV and the
  spectral index are the free parameters of the fit and are fitted in
  the energy range 500~MeV to 400~GeV.}
\label{table:morpho}
\end{table}

\begin{table}[htp!!]
\begin{minipage}{\linewidth}
  \renewcommand{\footnoterule}{}
\begin{tabular}{l|cccccc}
Source name & 
Flux\footnote{{\footnotesize{E$>$1~GeV, in $10^{-9}$~cm$^{-2}$~s$^{-1}$}}}&
Photon index &
Exp. cutoff \footnote{{\footnotesize{in GeV}}}&
~TS \footnote{{\footnotesize{Difference in TS value in comparison to a model
  with no source at the position of the respective source.}}} &
    R.~A. 2000&
    Dec. \\
\hline\hline
    1FGL J1705.5$-$4034c &  2.1 $\pm$  0.7  & 2.16 $\pm$  0.19  & &       20  \\
    1FGL J1709.7$-$4429 &  175 $\pm$ 6.4  &  1.74 $\pm$  0.03  & 4.46 $\pm$ 0.23 &    50064 \\
    1FGL J1714.5$-$3830c &  9.8 $\pm$  1.3  & 2.47 $\pm$  0.09  & &      228  \\
    1FGL J1716.9$-$3830c &  1.9 $\pm$  1.1 & 2.47 $\pm$  0.34  & &       14   \\
    1FGL J1717.9$-$3729c &  4.9 $\pm$  0.7  & 2.34 $\pm$  0.11  & &       81   \\
    1FGL J1718.2$-$3825 &  8.4 $\pm$ 4.3  &  1.64 $\pm$  0.41  & 1.72 $\pm$ 0.65 &  165 \\
    source {\emph{A}} &  1.6 $\pm$  0.5  & 2.03 $\pm$  0.17  & & 28 & 258.84$^{\circ}$ &$-$40.46$^{\circ}$ \\
    source {\emph{B}} &  4.2 $\pm$  1.2  & 2.48 $\pm$  0.16  & & 43 & 258.71$^{\circ}$& $-$38.70$^{\circ}$\\
    source {\emph{C}} &  2.5 $\pm$  0.7  & 2.45 $\pm$  0.22  & & 21 & 257.47$^{\circ}$ & $-$39.75$^{\circ}$ \\
    RX J1713.7$-$3946 &  2.8 $\pm$  0.7  & 1.50 $\pm$  0.11  & &       77  \\
  \hline 
\end{tabular}
\end{minipage}
\caption{Results of the spectral analysis of the gamma-ray emission in
  the ROI centered at \rxj. The integral flux between 1 and 300 GeV and the
  spectral index are the free parameters of the fit and are fitted in
  the energy range 500~MeV to 400~GeV.} 
\label{table:plawfit}
\end{table}

\newpage

\bibliographystyle{apj}
\bibliography{apj-jour,funk0203}

\end{document}